\documentclass[journal=jacsat,manuscript=article,layout=onecolumn]{achemso}
\usepackage{graphicx}
\usepackage{multicol}

\usepackage[version=3]{mhchem} 
\usepackage{xcolor}
\usepackage{parskip}
\usepackage{hyperref}

\definecolor{HarvardRed}{cmyk}{0.10,1,0.84,0.47}

\usepackage{lineno}

\author{Jelena Wohlwend}
\email{jelena.wohlwend@mat.ethz.ch}
 \affiliation{Laboratory for Nanometallurgy, Department of Materials, ETH Zurich, 8093 Zürich, Switzerland}
 \author{Georg Haberfehlner}%
 \affiliation{Institut für Elektronenmikroskopie und Nanoanalytik, TU Graz, 8010 Graz}
 \author{Henning Galinski}%
 \affiliation{Laboratory for Nanometallurgy, Department of Materials, ETH Zurich, 8093 Zürich, Switzerland}
\title[An \textsf{achemso} demo]
  {\textbf{Sequential Self-Assembly for Scalable Fabrication of Disordered Two-Phase Metamaterials}}

\abbreviations{IR,visible}

\begin{document}




\begin{abstract}
Self-assembly processes provide the means to achieve scalable and versatile metamaterials by "bottom-up" fabrication. Despite their enormous potential, especially as a platform for energy materials, self-assembled metamaterials are often limited to single phase systems, and complex multi-phase metamaterials have scarcely been explored. We propose a new approach based on sequential self-assembly that enables the formation of a two-phase metamaterial composed of a disordered network metamaterial with embedded nanoparticles. Taking advantage of both the high-spatial and high-energy resolution of electron energy loss spectroscopy, we observe inhomogeneous localization of light in the network, concurrent with dipolar and higher-order localized surface plasmon modes in the nanoparticles. Moreover, we demonstrate that the coupling strength deviates from the interaction of two classical dipoles when entering the strong coupling regime. The observed energy exchange between two phases in this complex metamaterial, realized solely through self-assembly, implies the possibility to exploit these disordered systems for plasmon-enhanced catalysis. 
\end{abstract}

\section{\small Introduction}

Plasmonic metamaterials and metasurfaces have attracted tremendous research interest in the field of photocatalysis and artificial photosynthesis because high absorption and local field enhancement in these plasmonic systems offer the potential to efficiently drive chemical processes by sunlight~\cite{solarenergy_review_plasmonics,swearerdionne_plasmoncatalysis,catalyticmetasurfacemaier,teri_plasmoniccatalysis,plasmoniccatalysis_nordlanderhalas}.
\par
To this effect, the damping of plasmons by non-radiative decay into so-called ”hot” carriers, i.e. electron-hole (e-h) pairs, has sparked considerable and long lasting theoretical and experimental interest. Especially, antenna-reactors complexes, often made of nanoparticle dimers, utilize hot carriers generated in the plasmon-enhanced near field to enhance chemical reactions by light~\cite{bimetallicnanostructuresplasmonicsandcatalysisdionne,antennareactor}.
\par
However, to guarantee technological impact these concepts must find a scalable analog.
\par
Among the strategies to design large scale nanostructured materials, self-assembly is a fundamental design principle ubiquitous in living organism~\cite{doi:10.1098/rsif.2014.1383}. Intriguingly, architectures created by self-assembly can be attributed to a single physical principle, the minimization of free energy in a nonuniform system. This principle is very general and applies to all length scales and material classes~\cite{Whitesides2002}.
\par
In fact, self-assembly has shown to be an effective approach to design large scale plasmonic metasurfaces~\cite{plasmonicselfassembly} for structural color displays using thin-film growth mechanisms~\cite{selfassemblyplasmonical}, to assemble plasmonic metamaterials by exploiting fluid instabilities~\cite{selfassemblyplasmonicmetamaterialsannealing} and to create fractal metamaterials for sensing applications based on self-assembly of hot aerosols~\cite{tricoli2020}.   
Despite the richness self-assembly offers for materials design, its full-potential including sequential combination of different self-assembly processes remains to be exploited.
\begin{figure*}[t!]
\includegraphics[width=1\textwidth]{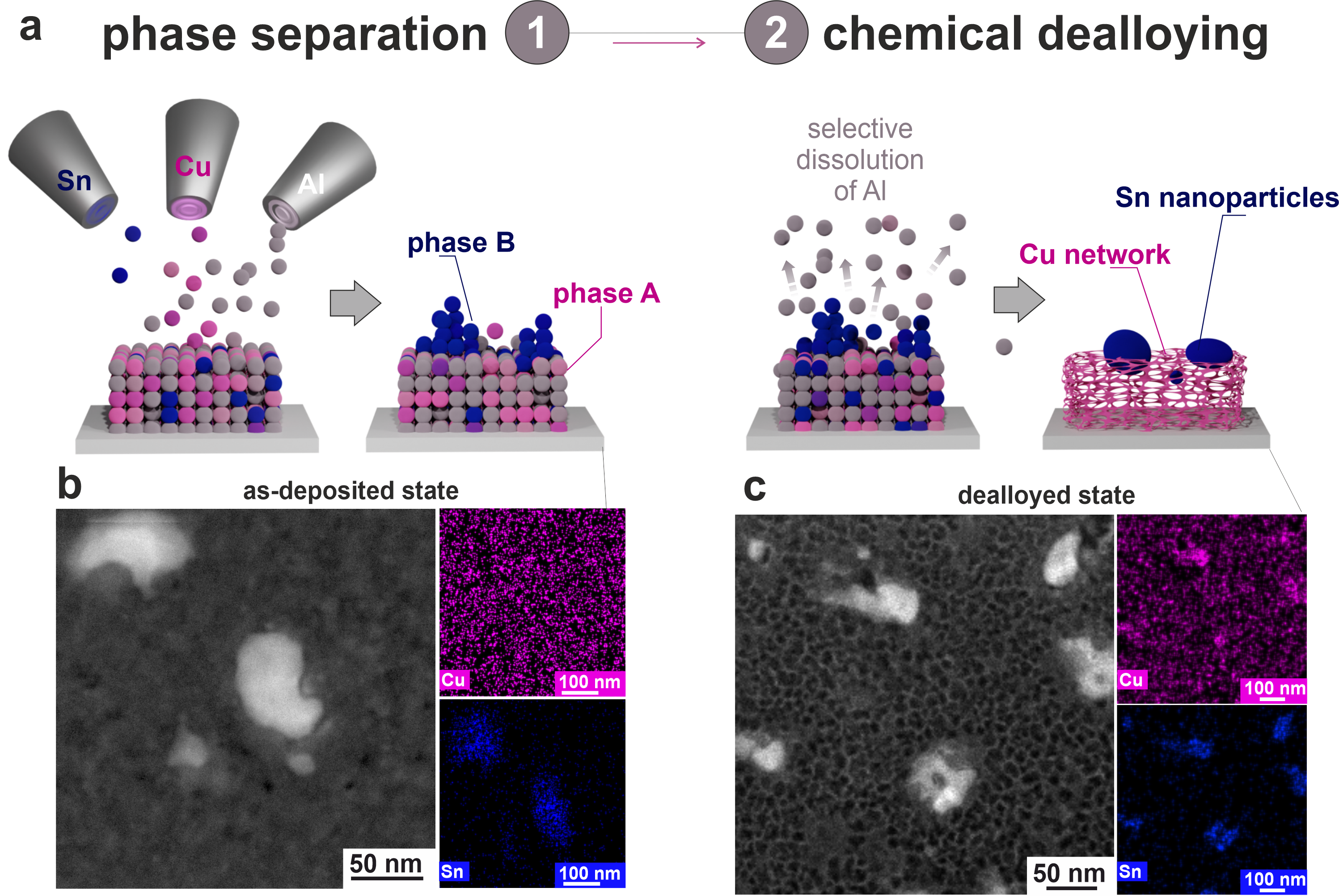}
\caption{\label{fig:one} \textbf{Sequential Self-Assembly of Disordered Two-phase Metamaterials.} \textbf{(a)} Schematic illustration of sequential self-assembly. First phase separation, meaning the Al-Cu-Sn thin films separate into phase A (Al + Cu) and phase B (Sn nanoparticles) during the co-sputter process of Al, Cu and Sn. Second chemical dealloying, where the ternary materials self-organiz into a Cu-based network and Sn nanoparticles by selective dissolution of Al. \textbf{(b)} High-angle annular dark field (HAADF) micrographs and elemental distribution maps of the as-deposited state confirming the phase separation of Al-Cu-Sn into a two-phase metamaterial, with phase A being the Cu-Al thin film and phase B being Sn NPs, during deposition. \textbf{(c)} HAADF micrographs and elemental distribution maps of the dealloyed state showing a Cu based disordered network with randomly distributed Sn NPs.}
\end{figure*}

\par
Here, we propose sequential self-assembly (SSA) as an adaptable and versatile platform to fabricate two-phase metamaterials (TPMs), i.e., metamaterials with at least two functional components. We select phase separation~\cite{phaseseparation_dufresne,phaseseparation_duringdepo_2,freeenergy} in an immiscible system and self-assembly by chemical dealloying~\cite{finely_divided_nickel,galinski2011dealloying} to realize our two-phase metamaterial (Figure \ref{fig:one}a). While the assembly by phase separation is due to the spontanous formation of domains from an unstable mixture~\cite{Kostorz1995}, self-organization in chemical dealloying results from a reaction-diffusion process, where the less noble metal in a solid solution is dissolved at the liquid-solid interface~\cite{nanoporousmetaldealloying}.

\section{\small Results}
\subsection{\small Sequential Self-Assembly of Disordered Two-phase Metamaterials}
\begin{figure*}[t!]
\includegraphics[width=1\textwidth]{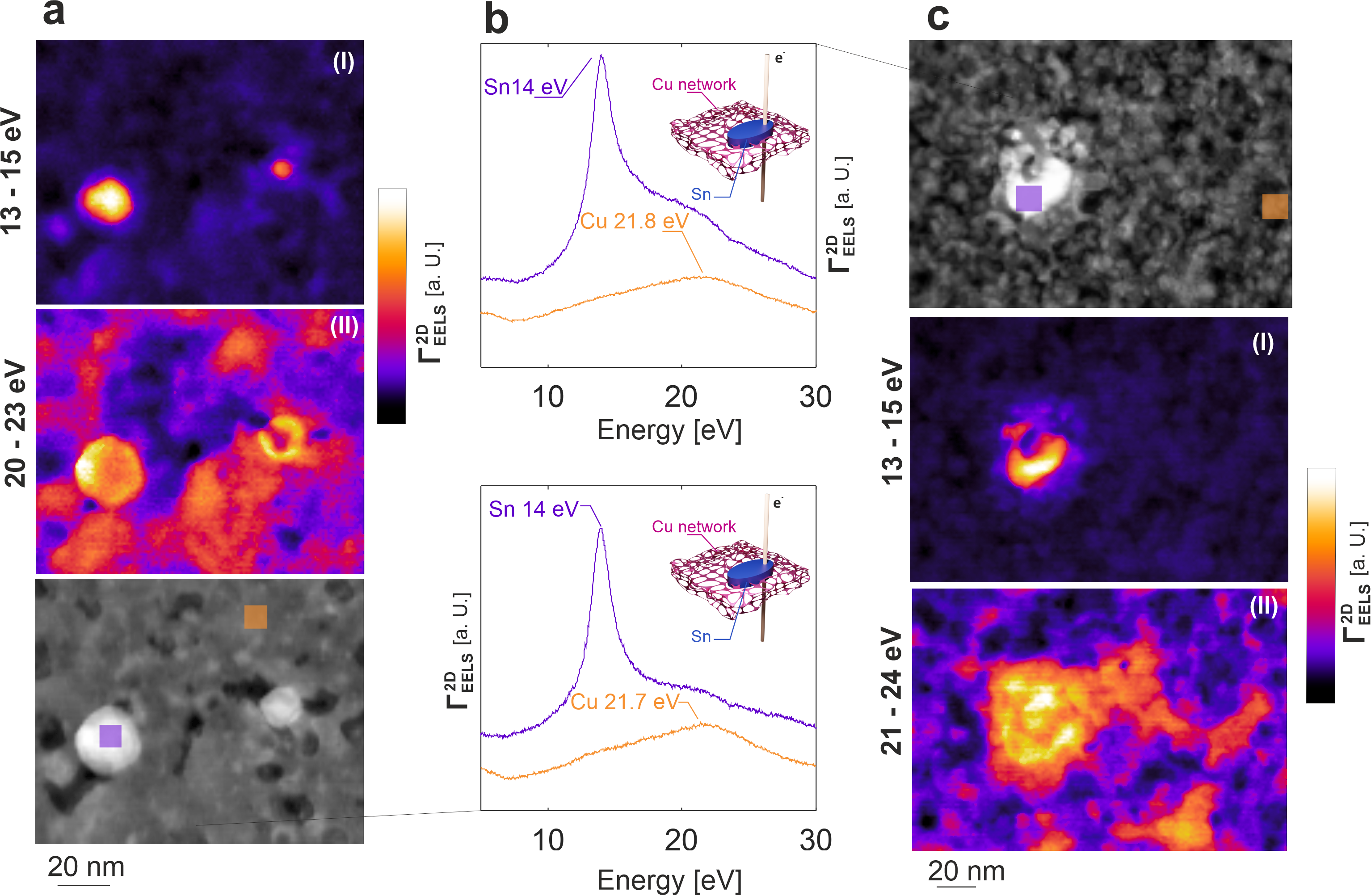}
\caption{\label{fig:two} \textbf{Chemical Analysis by Mapping Bulk Plasmon Modes} \textbf{(a)} and \textbf{(c)} EELS analysis of two metamaterials including HAADF micrographs of Cu–Sn networks with embedded Sn NPs, color-coded maps of bulk plasmon excitation of Sn NPs (I) and Cu-Sn network (II).\textbf{(b)} The specific regions of the spectra are indicated with colored squares in the HAADF micrographs. A clear shift of $\Delta E=$8 eV between the two bulk plasmon peaks is observed, confirming the formation of a two-phase system.}
\end{figure*}
Magnetron sputtering is ideally suited to generate conditions that facilitate phase separation on the nanoscale during growth~\cite{phaseseparation_duringdepo_1,phaseseparation_duringdepo_2}. Especially in immiscible co-sputtered systems, the interplay between the kinetics, such as the mobility difference of different elements, and the local thermodynamic potential favour phase separation. In case of binary Cu-Mo thin films the phase separation can be controlled by the substrate temperature, resulting in self-assembly of either vertically-, laterally-, or randomly- organized phases~\cite{ankit_pvd}. We transplant this approach to the ternary Al-Cu-Sn system which in the bulk phase diagram features a prominent miscibility gap (see also Supplementary Information)~\cite{alcusnphasediagramm}. Specifically, the deposition conditions were adjusted to favor growth kinetics that result in the formation of two phases at the nanoscale, visible in Figure \ref{fig:one}b.
\begin{figure*}[t!]
\includegraphics[width=1\textwidth]{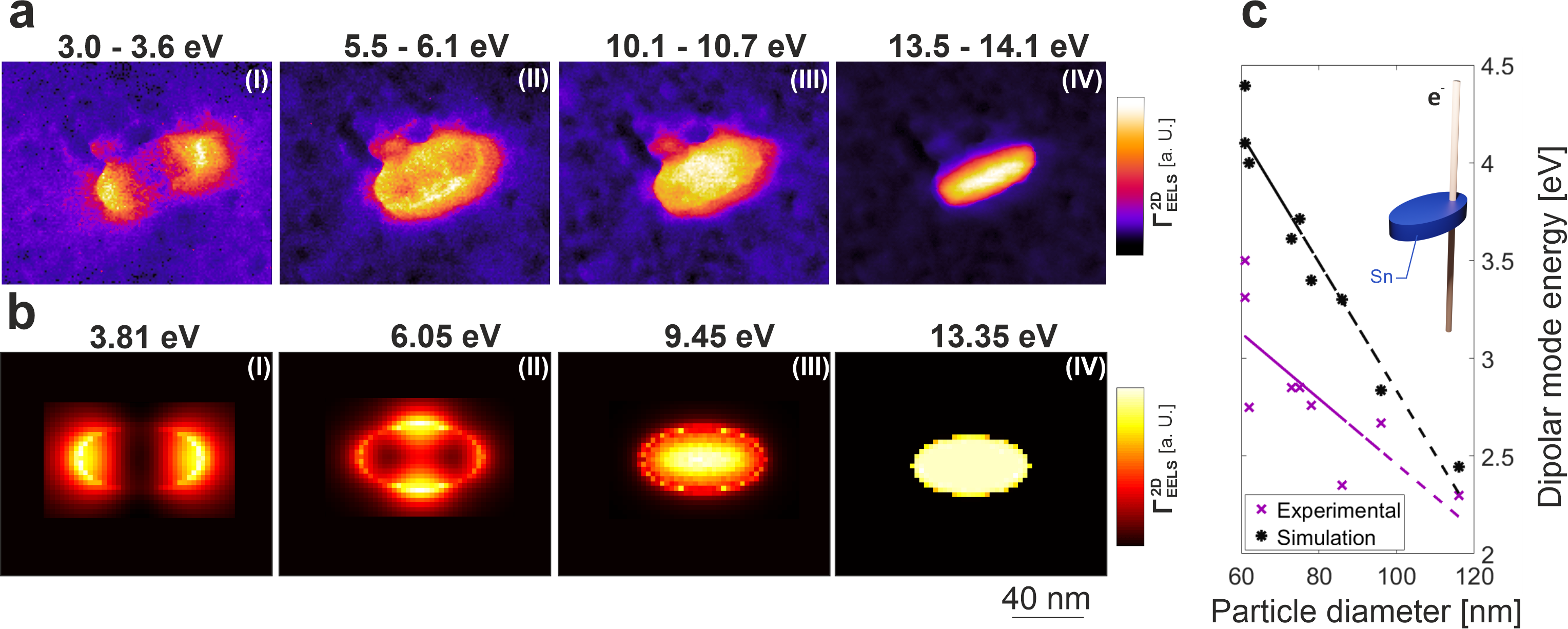}
\caption{\textbf{Localized Surface Plasmon Resonances (LSPR) in Sn Nanoparticles} \textbf{(a)} \& \textbf{(b)} measured and simulated  EELS maps for selected energies corresponding to the intensity maxima identified in the EELS spectra with \textbf{(I)} dipolar mode, \textbf{(II)} quadrupolar mode, \textbf{(III)} breathing mode and \textbf{(IV)} bulk plasmon mode. \textbf{(c)} Simulated and measured dependence of the dipolar mode energy on the long semi-axis of the Sn nanoparticle.}
\label{fig:three}
\end{figure*}
\par
After the initial phase separation, we make use of a second self-assembly process: chemical dealloying. Dealloying is the selective removal of the less noble alloy component to form an open-porous disordered network by a linearly propagating diffusion front~\cite{galinski2011dealloying}. Here, the majority element Al is removed in the dealloying process. The resulting self-organized two-phase system, consisting of a Cu network with embedded Sn nanoparticles (NPs), is shown in Figure~\ref{fig:one}c. 
\par
Elemental distribution maps of Cu and Sn in Figure~\ref{fig:one}b and c indicate the formation of two phases, i. e., Sn NPs and a Al-Cu thin film respectively a Cu network, through the SSA process (see also Supplementary Information).
\subsection{\small Chemical Analysis by Mapping Bulk Plasmon Modes}
To study the formation of these two-phase systems and to map their local plasmonic environment, we use electron energy loss spectroscopy (EELS). EELS has emerged as the ideal technique to characterize hybrid metamaterials on the nanoscale~\cite{kociakmappingplsamonsnanometerscale,tomographiplasmonicnanoparticle,modecouplingheterodimer,aluminumcayleytrees}. EELS is able to to excite and spatially probe surface and bulk plasmons. Thereby, chemical information can be obtained from the bulk plasmon peak of plasmonic nanostructures as the energy of the bulk plasmon is directly proportional to the electron density. 
\par
We study the local evolution of the bulk plasmon mode by EELS (Figure~\ref{fig:two}), to confirm the formation of a two phase-system. To underline the robustness and reproducibility of our SSA approach, two different samples are shown. The bulk plasmon energy $E_{\text{bulk}}=\hbar \sqrt{\frac{n_e e^2}{\epsilon_0 m_e}}$ scales with the free electron density $n_e$, and therefore is well suited to identify local chemical variations including phase formation at the nanoscale (Figure~\ref{fig:two}). To this extent, two different phases can be identified in the EELS maps and spectra shown in Figure~\ref{fig:two}: the Cu-based nanometric network with  $\omega_{\text{p}}=21.7$~eV and Sn nanoparticles with their bulk plasmon peak centered at $\omega_{\text{p}}=14.0$~eV.
The measured bulk plasmon energy of Sn is in good agreement with values reported in literature~\cite{snsmallmetallicsphereseels,EELSAtlas}, whereas the Cu peak slightly deviates from reported values~\cite{Cu_CuO,EELSAtlas}. This deviation might stem from residual Al or partial oxidation shifting the bulk plasmon peak to higher energies. Remarkably, both size and shape of the nanoparticle phase can be engineered by changing the kinetics of phase separation during growth (see also Supplementary Information), so that nanoparticles with spherical (Figure~\ref{fig:two}a), ellipsoidal (Figure~\ref{fig:three}a) and arbitrary shapes (Figure~\ref{fig:one}b and c) can been realized.
\begin{figure*}[h!]
\includegraphics[width=1\textwidth]{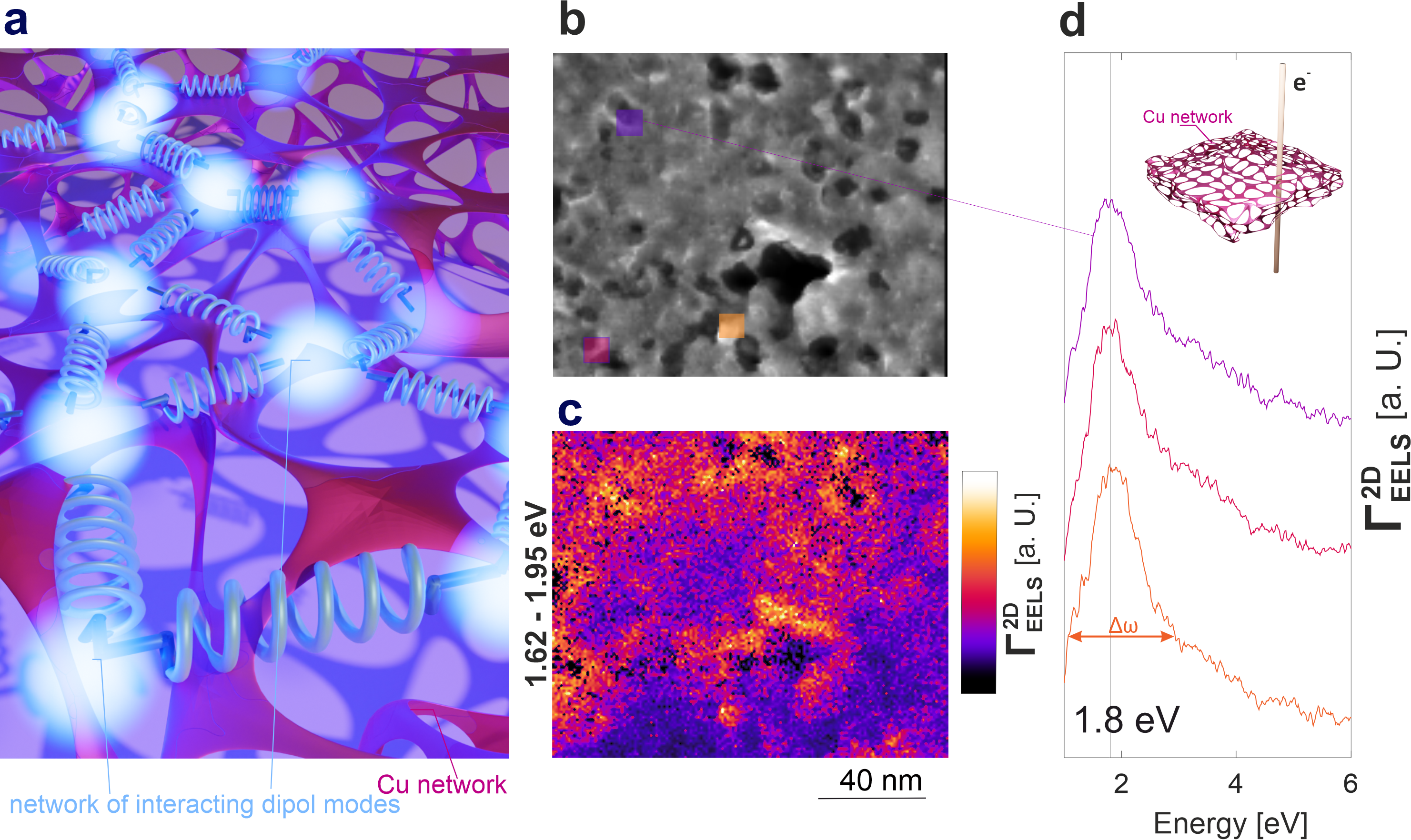}
\caption{ \textbf{Global Network Mode in Disordered Network Metamaterials} \textbf{(a)} Schematic illustration of the plasmonic response in a network metamaterial, exhibiting a set of coupled electromagnetic dipole-like resonances (balls \& springs), whose distribution and interaction is given by the local connectivity and curvature of the dealloyed Cu-network. \textbf{(b)} HAADF image of a disordered Cu network metamaterial. \textbf{(c)} Color coded EELS map containing global network modes (GNMs) with an energy of ~1.8 eV forming local hot spots on network studs and \textbf{(d)} EELS spectra from different hot spot regions, $10\times10$~nm$^2$, highlighting the efficient localization of light  over a broad frequency range$(\Delta\omega=1.8 \pm 0.076~eV)$. The specific regions are indicated with colored squares in (b).}
\label{fig:four}
\end{figure*}
\subsection{\small Localized Surface Plasmon Resonances in Sn Nanoparticles}
We begin to study the complex photonic environment of the two-phase network metamaterial, by characterizing the modal properties of  the nanoparticle phase. The sustained localized surface plasmon (LSP) modes of an ellipsoidal Sn NP excited by the electron beam are reported in Figure~\ref{fig:three}. To achieve further insight on the characteristics of this modes, the EELS signal is mapped within a selected energy range centered around the resonance energy of each mode (Figure~\ref{fig:three}a~I-IV).  The Sn NP exhibits several localized SP resonances including a dipolar mode centered at 3.3 eV and higher order (HO) modes including a quadrupolar mode (5.8 eV) and a breathing mode~\cite{darkplasmonicbreathingmode} (10.4 eV). Quadrupolar modes and breathing modes are typically optically dark excitations, meaning they cannot be excited by light, highlighting again the advantages of characterizing plasmonic structures by EELS~\cite{darkplasmonicbreathingmode}. 
The bulk plasmon peak in Figure~\ref{fig:three}a~IV visualizes the spatial dimensions of the Sn NP and additionally confirms the chemical nature of this particle~\cite{snsmallmetallicsphereseels}.
\par
Figure~\ref{fig:three}b compares the experimentally obtained EELS data with electron energy-loss simulations performed using the MNPBEM toolbox~\cite{Matlab_toolbox}. Here, the Sn nanoparticle is modeled as an elliptical cylinder with a height of 15~nm using refractive index data from literature~\cite{plasmonicsinthuv}. Impressively, the obtained resonance energy and simulated EELS maps shown in Figure~\ref{fig:three}b are in good agreement with the experiment. While all experimentally observed modes are reproduced, slight deviations in the resonance energy are found. These deviations can be explained by the influence of the surrounding Cu-network on the non-retarded resonance energy of the modes~\cite{raza2015multipole} as well as by deviations of the experimental NPs from the simulated elliptical cylinder.
\par
The LSP modes of metallic nanoparicles are highly dependent on shape and size, a trend also observed in our Sn NPs~\cite{modecouplingheterodimer,wheredoesenergygoineels}. With an increase in particle diameter (long semi-axis) the dipolar mode is blue shifted, as shown in Figure \ref{fig:three}c, where the dipolar mode energy is displayed as a function of the particle diameter. A similar linear scaling is found for simulated elliptical Sn nanoparticles (Figure \ref{fig:three}c), while the difference in slope might originate again from the surrounding medium as well as from size and shape differences between the simulated elliptical cylinders and the experimentally measured imperfect Sn NPs~\cite{aberrantgoldnanostructures}.  
\subsection{\small Global Network Mode in Disordered Network Metamaterials}
Next, we characterize the plasmonic response of the second phase: the Cu-based dealloyed network. These networks can be represented as coupled dipole-like networks (Figure~\ref{fig:four}~a), where the disordered network traps and localizes surface plasmon (SP) waves~\cite{galinski}. Figure \ref{fig:four}~b shows a HAADF scanning transmission electron micrograph of such a disordered network. The disorder, which is a product of self-organization during dealloying, is characterized by the complete absence of translational symmetry and an abundant variety of different curvatures that arise from local variation in edge length and connectivity (see also Figure~\ref{fig:one}~c). To visualize the distribution of plasmonic modes of the network, we map the EELS signal, centered at $1.8$~eV within a spectral window of $0.33$~eV (Figure~\ref{fig:four}~c). We observe large fluctuations in EELS intensity (Figure~\ref{fig:four}~c) which is linked to the local density of optical states (LDOS)~\cite{jelena2022}. Such fluctuations confirm both the highly inhomogeneous spatial distribution of the plasmonic eigenmodes and their ability to localize energy at nanometric scales in so-called "hot-spots", i. e. regions of high field intensities~\cite{Carminati2010}. 
\par
To demonstrate the effect of disorder on the localization in the network, we probe three different hot-spots with a fixed area of $10 \time 10$~nm$^2$ (Figure~\ref{fig:four}~b) and compare the corresponding EELS spectra in Figure~\ref{fig:four}~d. Remarkably, all three hot-spots localize polychromatic light very effectively while their spectra have quasi identical line-shape featuring the same broadband response $\Delta\omega= 1.80 \pm 0.07$~eV centered at 1.8~eV. We can infer the formation of such a broadband plasmonic response to the disorder in the system, causing an equal distribution of the input energy between all plasmonic modes of the network~\cite{galinski,manipulatingdisorder}. Here, the localization length of these modes appears large enough, so they can extent over several "hot-spots"\cite{Stockman1994}, resulting in a delocalized "global" plasmonic response of the network. Reminiscent to connected wire-mesh metamaterials~\cite{metamaterialspendry}, these global network modes (GNMs) rather rely on the connectivity of the network\cite{jelena2022,galinski}. 
\subsection{\small Mode Coupling}

\begin{figure*}[t!]
\includegraphics[width=0.78\textwidth]{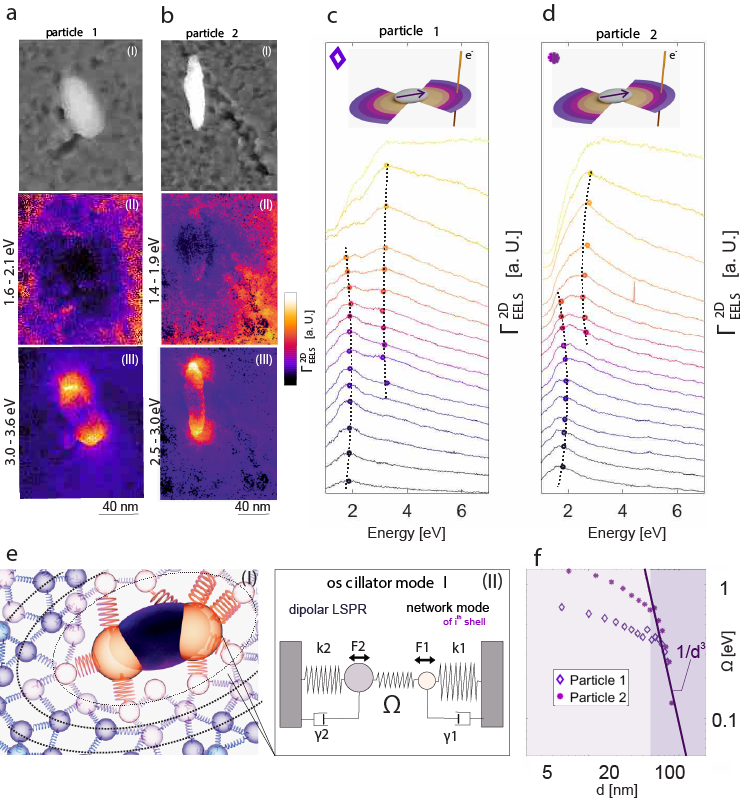}
\caption{ \textbf{Mode Coupling} \textbf{(a)} and \textbf{(b)} (I) HAADF image of disordered two-phase metamaterials with color-coded maps of network mode (II) and dipolar mode (III). \textbf{(c)} and \textbf{(d)} waterfall plot of EEL spectra averaged in the shell sectors around nanoparticle 1 and 2, where the dipolar mode of the Sn NP is dominant. The color transition from yellow to orange to violet, corresponds to an increased in shell number around the NPs. Peak maximas (GNM and dipolar mode) from fitting the oscillator model to the experimental data are indicated by colored dots in the EEL spectra. The peak shift of both modes is indicated by a dotted black line. The figure inset shows schematically the division of the EELS maps into sector shells around the NP \textbf{(e)} (I) Schematic illustration of the coupling between the GNM and the dipolar nanoparticle mode with the first shell around the NP shaded in pink. (II) Schematic illustration of the oscillator model, i. e. two forced damped coupled harmonic oscillators. \textbf{(f)}  Calculated coupling constant $\Omega$ for particle 1 (purple diamond) and particle 2 (pink star) as a function of the distance $d$ to the Sn NP. The two different coupling regimes are shaded in purple, weak coupling with $\Omega$ scaling with $1/d^3$ and pink indicating strong coupling in close proximity to the nanoparticle.}
\label{fig:five}
\end{figure*}
To study the interaction of the network modes and LSP modes, we investigate the EELS probability in the vicinity of the nanoparticles. Figure \ref{fig:five} a and b II each show the EELS intensity of network modes in the vicinity of a single ellipsoidal Sn-nanoparticel. The signal is strongly modulated and radially decreasing around the NP, suggesting that the mode spectrum is significantly modified and energy is exchanged.
\par
To describe this mode coupling and its radial dependence in more detail, it is convenient to divide the area around the NP in thin nanometric shells. To only consider the high field regions of the dipolar mode of the Sn NPs (Figure \ref{fig:three}a I), the shells are further divided into four sectors using the orientation (long axis) of the NPs. A more detailed approach to the segmentation of the EELS maps is found in the Supplementary Information. 
\par
By calculating the average EELS spectrum in each shell, we retrieve the evolution of the EELS spectra at discrete distances around the nanoparticle (Figure~\ref{fig:five}c and Figure~\ref{fig:five}d). Here, we limit ourselves to sectors in direction of the long axis of the Sn NPs, while the other direction is given in the Supplementary Information. In the waterfall plots in Figure~\ref{fig:five}c and Figure~\ref{fig:five}d, we find that the spectra of the GNM ($\approx 1.8$eV) and the dipolar LSP($\approx 3$eV) clearly shift when moving away from the nanoparticle center. The dashed trend line in Figure \ref{fig:five}c and d suggests avoided crossing~\cite{avoidedcrossing} of the GNM and the dipolar mode, a phenomena where two modes first approach and then repel each other. Avoided crossing is evocative of mode coupling.
\par
To paint a fuller physical picture, we use a set of coupled damped harmonic oscillators to model the mode coupling~\cite{nanoantennasengineeringinstacked}. To simplify, we disregard  inter-shell interactions and only consider interaction of one effective network mode with the dipolar LSP mode within a single shell (Figure~\ref{fig:five}e~I).   
\par
In the quasi-static limit, the two driven coupled oscillators (Figure~\ref{fig:five}e~II) located at positions $r_i$ and driven by the force $F(r_i,t)$ are described by the effective Hamiltonian
\begin{equation}
\scriptsize
    H= \sum_{i=1} ^{N} \frac{P_i^2}{2 m_i} + \frac{1}{2}m_i \omega_i^2 Q_i^2 -\sum_{i\neq j} Q_i \Omega_{ij} Q_j
    - \sum_{i=1} ^{N} F(r_i,t) Q_i,
\end{equation}
where $Q_{i}$ and $P_{i}$ represent the displacement coordinate and conjugate momentum, $m_i$ the effective mass, and $\omega_i$ the resonance frequency of the $i$th dipolar mode. Furthermore, $\Omega_{ij}$ is the coupling strength for pairwise near-field dipole-dipole interaction. A detailed discussion on the model solution and shell fitting can be found in the Supplementary Information. We find good agreement between the experimental data and our model, in terms of peak width and spectral position. The maxima of the modeled spectra are indicated by colored dots in Figure~\ref{fig:five}c and d and connected by a dashed trend line.
\par 
Moreover, the coupled oscillator model can be used to determine the coupling strength $\Omega_{ij}$ between the global network mode and the dipolar LSP mode in each cell. The fitted coupling constant $\Omega$ as a function of $d$ is shown in Figure~\ref{fig:five}f for the two selected particles.  
\par
In case of two classically interacting dipoles, we expect $\Omega$ to follow $1/d^3$ in the near-field~\cite{interparticlecoupling}. We observe this scaling for the outer shells, visible in Figure~\ref{fig:five}f, shaded in light purple. In this region $\Omega < \omega_i$ the interaction lie in the weak coupling regime. An important finding of our analysis is observed with $d$ becoming smaller ($d<50nm$), where $\Omega \approx \omega_i$ stops to follow $1/d^3$ (Figure~\ref{fig:five}f, shaded in light pink) and is accompanied by significant peak broadening (Figure \ref{fig:five}c and d). This finding indicates a change in the coupling regime from weak to strong coupling in close proximity to the Sn NPs, as $\Omega$ approaches $\omega_i$. We therefore find striking similarities between our system and the break down of dipol-dipol interaction in NP dimers with ultra small gaps~\cite{interparticlecoupling}. The observed mode coupling outlines the possibility of energy exchange between the disordered network metamaterial and a second system over a large area. Realizing energy exchange, i.e. mode coupling in novel materials based on self-assembly offer pathways towards large scale energy materials for plasmon-enhanced catalysis.
\section{\small Discussion}
In conclusion, we have experimentally demonstrated a general principle based on sequential self-assembly for fabricating large-area two-phase metamaterials. By selecting two specific self-assembly reactions, namely phase separation and chemical dealloying, a two-phase metamaterial consisting of a disordered Cu network metamaterial and Sn nanoparticles has been manufactured. Engineering of the size and shape of the nanoparticles and the network~\cite{jelena2022} is readily achieved by altering the deposition parameters. As our approach relies on basic thermodynamic principles, it may apply to other ternary systems with a miscibility gap and also to systems with the tendency to form metastable intermetallics.
\par
Electron energy loss spectroscopy allowed for fundamental insight in the modal properties of these two-phase metamaterials. We presented first experiments showing  inhomogeneous localization of optical energy in "hot spots" of the network. The spectra of plasmonic eigenmodes measured in different "hot-spots" of the network are quasi identical, suggesting that the network provides a means for energy equipartition. Previous experimental works~\cite{jelena2022,galinski} studying similar networks in the context of perfect absorbers, imply that a significant fraction of these plasmonic eigenmodes are accessible from the far field.
\par
Beyond the localization of light in the network phase, we demonstrate that mode-coupling is achieved between the plasmonic modes of the network and the nanoparticles. In close proximity to the nanoparticle, the system transitions to the strong coupling regime which goes alongside with a breakdown of the classical dipole-dipole interaction. This presents a promising approach for designing a variety of coupled two-phase metamaterials which may include a dielectric phase or active phase, such as quantum dots, in addition to the plasmonic phase.   
\par
The possibility to design complex coupled systems solely by self-assembly and the ability to effectively localize polychromatic light at the nanoscale, offer unique opportunities to exploit these two-phase metamaterials as a platform for light harvesting and plasmon-enhanced chemistry. 
\section{\small Materials and Methods}
\par  \textbf{Sample fabrication.} The samples were created by a two step process. In a first step, Al-Cu-Sn thin films were co-sputtered onto commercial carbon supported gold TEM-grids. In a second step, the films were chemically dealloyed in a 1 M NaOH aqueous solution, to form a disordered open-porous nano-network.
\\ \\ \textbf{\small EELS Measurements.}
All EEL spectra were obtained by a monochromated FEI Titan 60-300 with an imaging filter (Gatan GIF Quantum) operated in scanning mode at 300 kV. The spectra were acquired with a dispersion of 0.25, respectively, 0.1 meV per pixel. Additionally, all spectra were treated with the HQ Dark Correction plugin.
\\ \\ \textbf{\small EELS Data Processing.}
Postprocessing of the EELS data included the alignment of the spectra with respect to the position of the zero-loss peak (ZLP), normalization of the maximum intensity of the ZLPs in each pixel and removal of the ZLP by fitting of a premeasured ZLP using the Matlab Spectrum image analysis tool~\cite{Matlab_toolbox} and DigitalMicrograph (Gatan). Additionally, a script for DigitalMicrograph was developed to analyse the EEL spectra in relation to the distance from a selected particle.
\\ \\ \textbf{\small Simulations.}
Complementary simulations were preformed using the MNPBEM toolbox with the Sn nanoparticles modeled as an extruded ellipse~\cite{Matlab_toolbox}. The dielectric function of Sn was approximated by a Drude-Lorentz model $\epsilon(\omega) \approx \epsilon_\infty - \frac{ {\omega_D}^2}{\omega ( \omega+ i \gamma_D) } - {\sum_{n=1}}^{N_L} \frac{  \epsilon_{Ln} {\omega_{Ln}}^2} {\omega ( \omega+2 i {\delta_{Ln}}^2) - {\omega_{Ln}}^2}$ whereby $\epsilon_\infty = 1.203$, $\hbar\omega_D =12.439$ (eV) and $\hbar\gamma_D =$0.757 (eV) taken from McMahon \textit{et. al}. \cite{plasmonicsinthuv} \par 
\textbf{Acknowledgement} \par 
The authors cordially thank the Graz Centre for Electron Microscopy for allowing access to their facilities. The authors acknowledge the technical support of FIRST as well as BRNC. The authors are also grateful to Nikolaus Porenta and Julian Gujer for their help and support. The authors thank Joan Sendra and Rebecca Gallivan for useful comments and careful reading of the manuscript. This project has been funded in part by the European Union’s Horizon 2020 research and innovation programme under grant agreement No 823717 – ESTEEM3.\\ \\
\textbf{Author contributions}\\ \\
J. W., H. G. and G. H. conceived the research plan.
J. W. fabricated the samples. J. W. and G. H. collected EELS maps. J. W., H. G. and G. H. performed the data analysis. J. W. and H. G. wrote the manuscript and visualized the data. All authors reviewed and commented the manuscript.\\ \\
\textbf{Competing interests}\\ \\
The authors declare no competing interests. 

\bibliography{paper.bib}


\end{document}






\subsection{The Cu-Sn-Al system}
\begin{figure*}[h!]
  \includegraphics[width=0.5\textwidth]{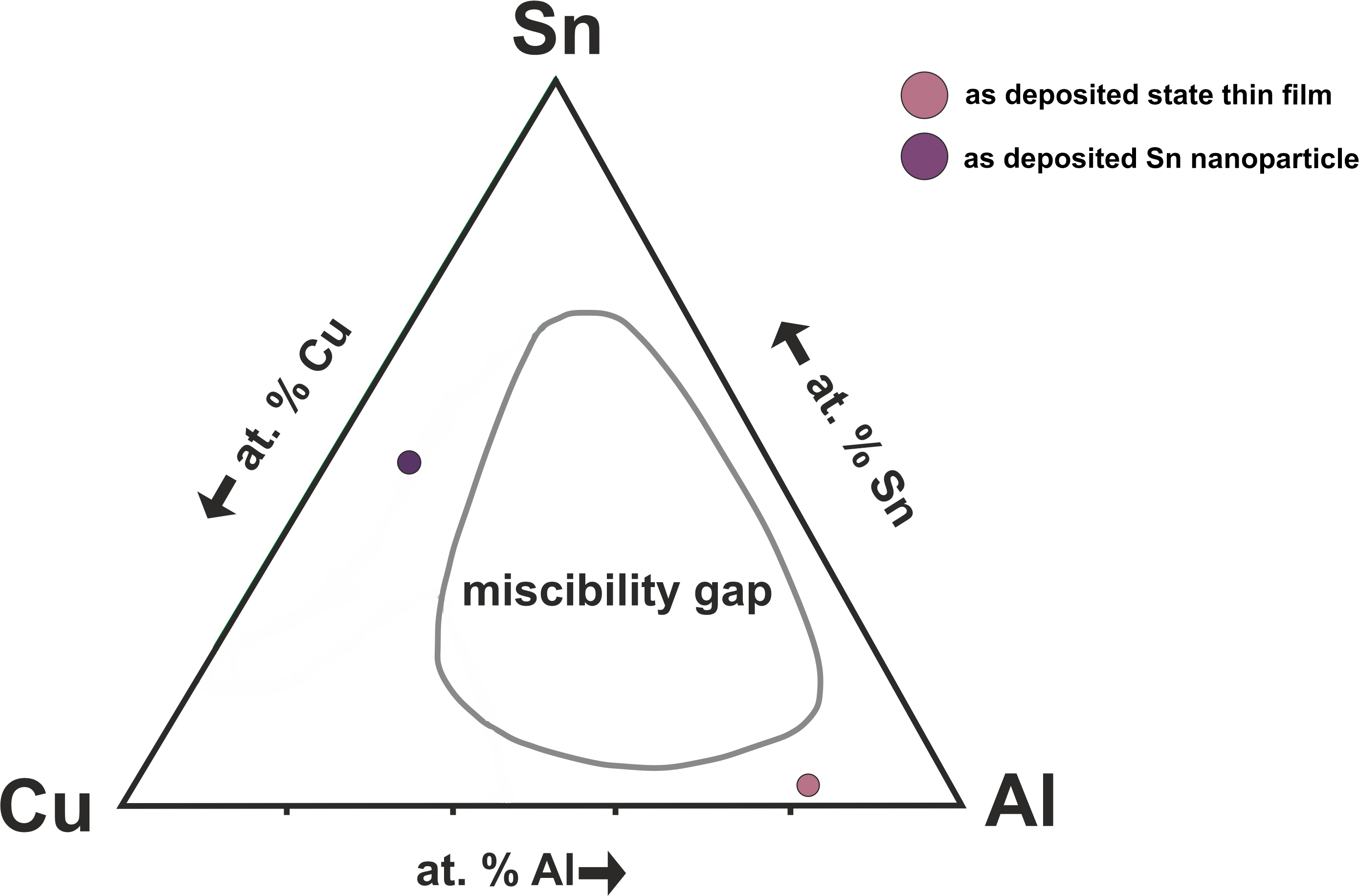}
\caption{Ternary Phase Diagram of Al-Cu-Sn with miscibility gap\cite{alcusnphasediagramm}
}
\label{fig:phasediagramm}
\end{figure*}

To form disordered metallic networks by chemical dealloying a minimum of two alloy components are required. Of these, one component must stay unoxidized while the other is dissolved in the electrolytic environment. Here we use Al, due to its high oxiphilicity, meaning it can be oxidized easily to e.g. sodium aluminates in sodium hydroxide solutions. The Al content in the as-deposited alloy must range between 60 and 80 at. \% as due to the parting limit surface passivisation occurs at lower percentages~\cite{dealloyingcu}. 
\\
\\
For the backbone of the disordered network we choose Cu as it is a low-loss alternative to silver for plasmonic applications~\cite{cuplasmonics}. . 
\\
\\
As a second phase we choose Sn as the Al-Cu-Sn phase diagram shows a miscibility gap over a wide range of compositions (see Figure \ref{fig:phasediagramm}). We outline the possibility to form phase-separated systems due to this miscibility gap already in previous work~\cite{wohlwend}. Additionally, Sn in combination with Cu is a promising material in regards to catalysis~\cite{cu_sncatalysis,SnCatalysisforElectrochemicalReductionofCO2}. 
\\
\\
\subsection{Phase separation during physical vapor deposition}
\begin{figure*}[h!]
  \includegraphics[width=1\textwidth]{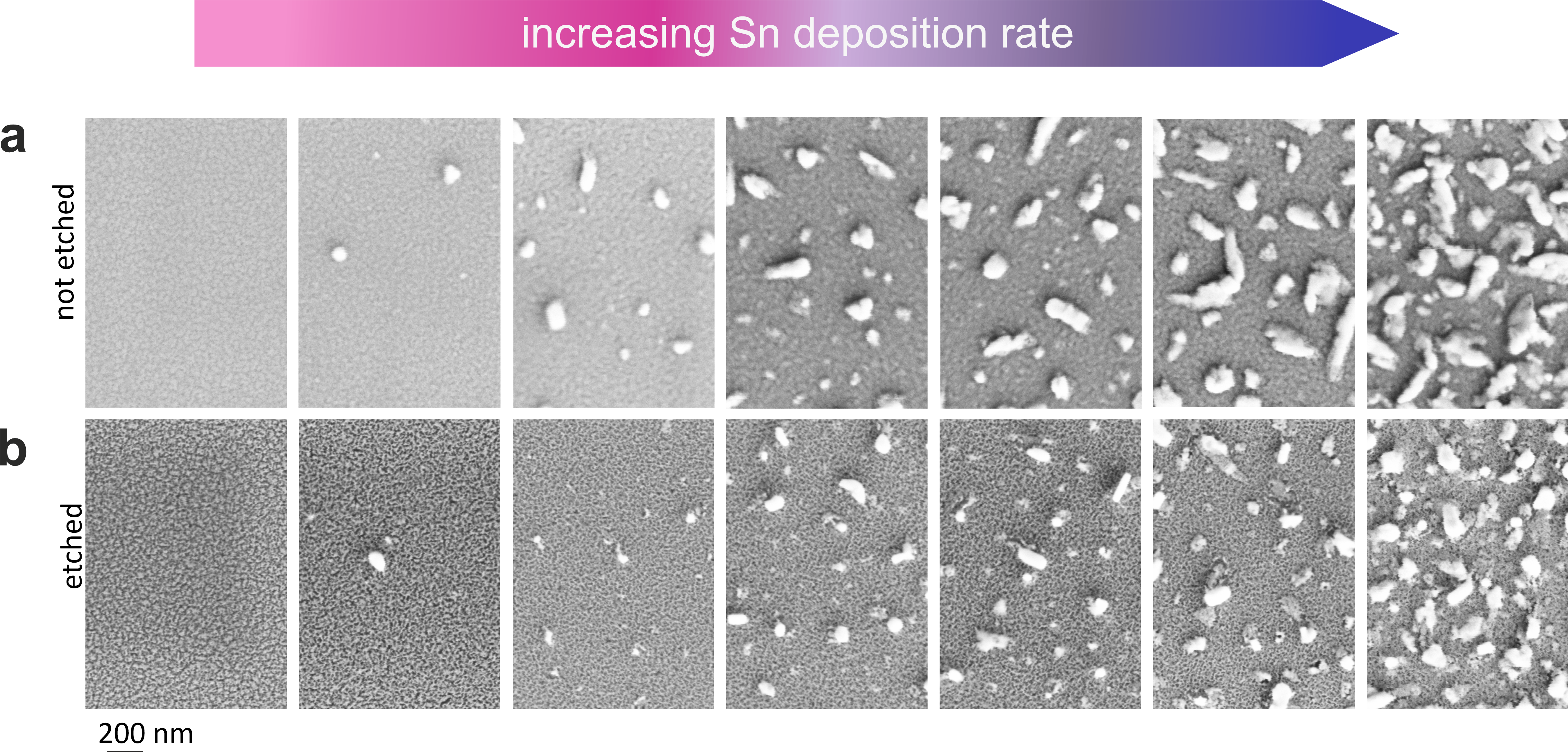}
\caption{\textbf{Nanoparticle formation depending on the Sn deposition rate} Scanning electron micrographs (top view of the films) of the as deposited films (a) and dealloyed films (b) with increasing Sn deposition rate $r_{sn}$ from left to right.}
\label{fig:figure_three_incresingrate}
\end{figure*}

Physical vapor deposition offers a high number of adaptable process parameters such as magnetron power, throw distance, bias voltage, base pressure, Ar pressure, substrate temperature, substrate selection and deposition rate for the fabrication of thin films~\cite{POWERScosputtering,pvd}. Additional parameters are introduced if multiple elements are deposited at the same time, i. e., co-sputtered. These parameters govern the thermodynamics and kinetics of the sputtered atoms as they land on the surface of the substrate. For immiscible co-sputtered systems, the interplay of mobility differences of different elements and phase separation forces drive the morphology and phase formation of the sputtered thin films. Through careful selection of the processing parameters phase-separated systems can be created in a multitude of forms~\cite{phaseseparation_duringdepo_1,phaseseparation_duringdepo_2}. One example of a designed phase-separated system are co-sputtered Cu-Mo thin films. With increasing substrate temperature the system transforms from a vertically organised (phase-separated) layered structure over a laterally organised layered structure to a randomly phase-separated structure~\cite{substratetempcumofilms}.
\\
\\
In our experiments, the temperature as well as the deposition rate of Al ($ r_{Al}=0.064~\frac{nm}{s}$) and Cu ($r_{cu}=0.025~\frac{nm}{s}$) were kept constant. The deposition rate $r_{sn}$ of Sn however, was varied. This results in the formation of Sn NPs in various sizes and shapes depending on $ r_{sn}$. In Figure~\ref{fig:figure_three_incresingrate} scanning electron micrographs of the as deposited and dealloyed films are shown for a increasing $r_{sn}$ of Sn from left to right. With increasing $r_{sn}$ of Sn a significant change in the particles on the film surface is observed. If the deposition rate is low, circular particles form and with increasing $r_{sn}$ the particles grow in size and irregularity, from circular over elliptic to arbitrary. 
\\
\\
The observed phase separation during the co-sputtering of Al-Cu-Sn most probably stems from the interplay of mobility differences and phase separation driving forces. This can be seen by looking at the average diffusion distance, described by the relation $\chi  = \sqrt{2 \nu \frac{a}{r}} a e^{\frac{-Q_s}{2kT}}$ with k being the Boltzmann constant, \textit{a} the individual jump distance depending on interplanar spacing, \textit{r} the deposition rate, \textit{T} the deposition temperature and $Q_s$ the activation energy for surface diffusion~\cite{POWERScosputtering}. Therefore, the average diffusion of the different elements can be influenced by changing the deposition rate and temperature. Intuitively, if the deposition rate is too high, there is no time for the atoms to rearrange before the next one arrives resulting in frozen kinetics. With lower deposition rates, however, the atoms have time to rearrange before another one arrives, consequently a multitude of phase-separated architectures might arise~\cite{phaseseparation_duringdepo_1,phaseseparation_duringdepo_2}. Compared to commonly used deposition rates, it has to be notes that our rates are at least an order of magnitudes lower~\cite{substratetempcumofilms}.
\subsection{Elemental composition}
\begin{figure*}[h!]
  \includegraphics[width=1\textwidth]{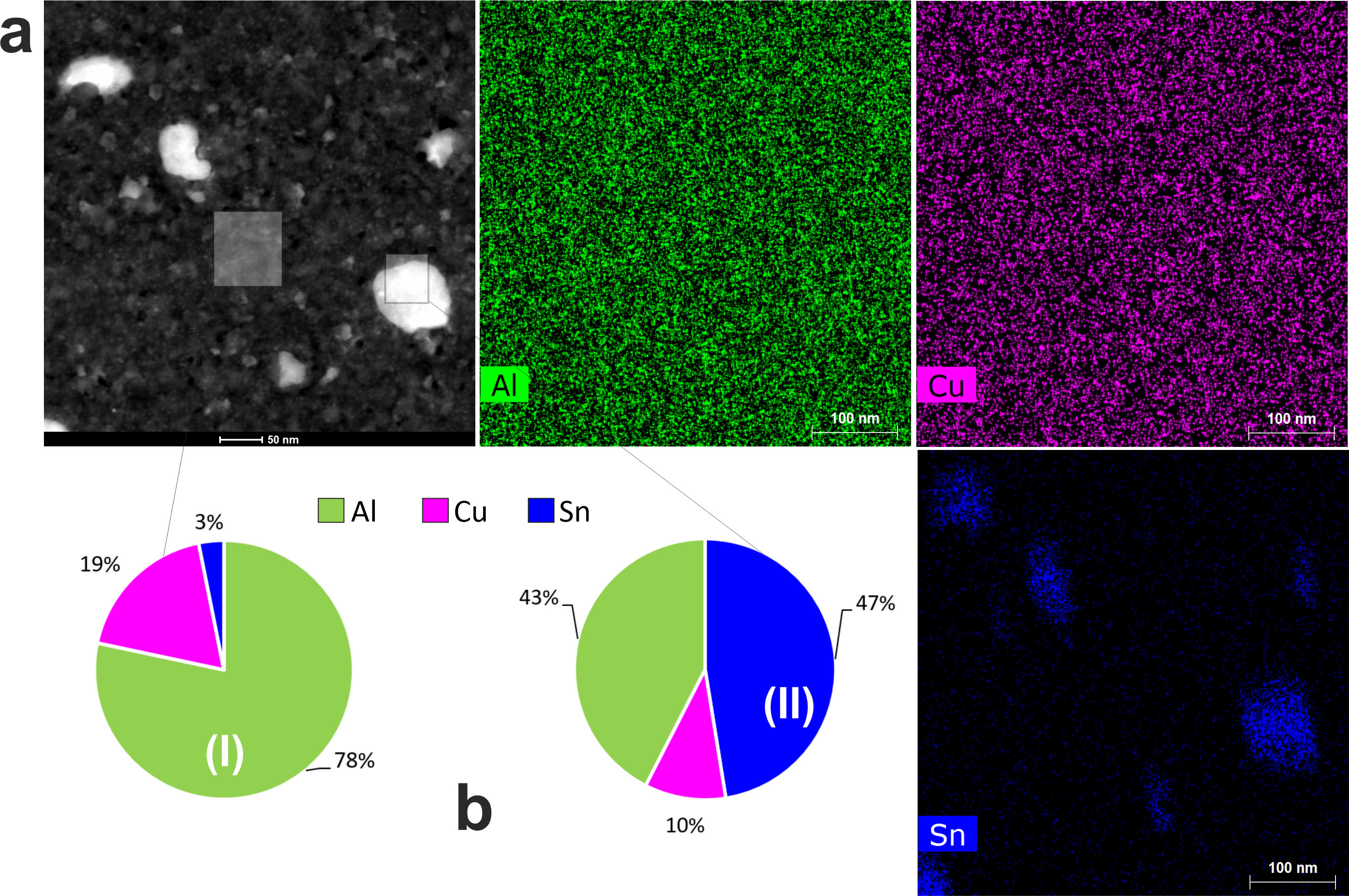}
    \caption{\textbf{Elemental composition of as deposited state} (a) Dark field and elemental content  distribution of the as deposited state with (b) cake diagrams illustrating the elemental content distribution of selected regions marked in the DF image
}
\label{fig:elementaldistribution}
\end{figure*}

\begin{figure*}[h!]
  \includegraphics[width=1\textwidth]{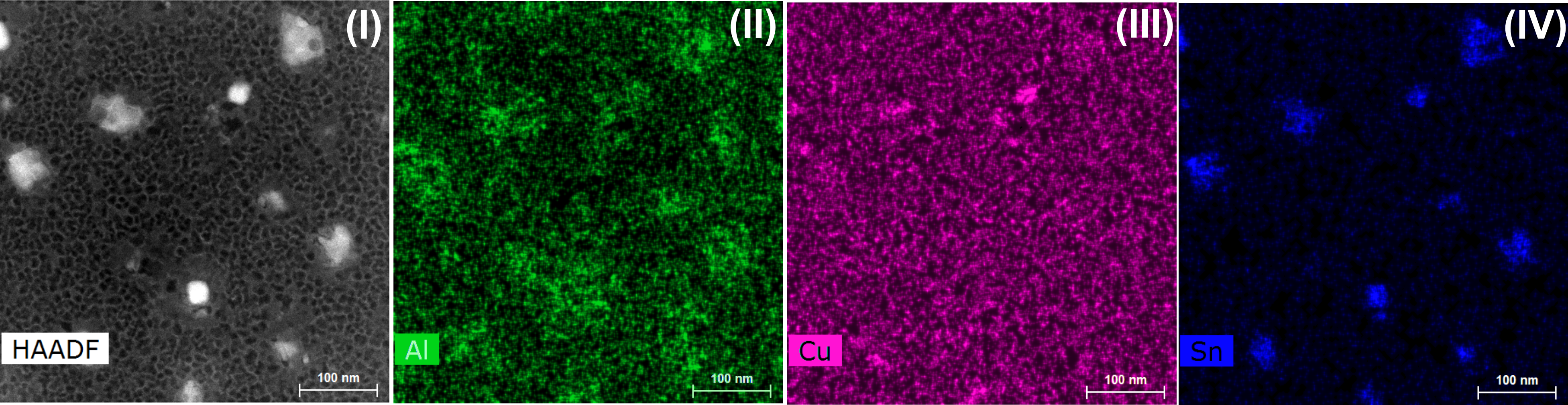}
    \caption{\textbf{Elemental maps of the dealloyed state}(I) High angle angular dark field image and elemental distribution map of (II) Al (III) Cu (IV) Sn}
\label{fig:elementaldistribution2}
\end{figure*}

Figure~\ref{fig:elementaldistribution} presents the elemental content distribution maps and the composition evaluations of the I thin film and II nanoparticle (NP) of the as deposited state. Al and Cu are uniformly distributed over the sample, whereas Sn is primary found in the NPs. As to be expected in the as-deposited state, Al is found to be the majority element.
In Figure~\ref{fig:elementaldistribution2} the elemental distribution maps of the etched state are shown. The nanomodulations, formed by the chemical dealloying process, of the disordered network metamaterial are clearly seen in the HAADF image and also observed in the elemental map of Cu (Figure \ref{fig:elementaldistribution} I and III). Some Al is still remaining after the dealloying process (Figure \ref{fig:elementaldistribution} II). In Figure \ref{fig:elementaldistribution} IV the after the dealloying process remaining Sn NP are clearly visible.
\subsection{EELS Map Segmentation}
\begin{figure*}[h!]
  \includegraphics[width=1\textwidth]{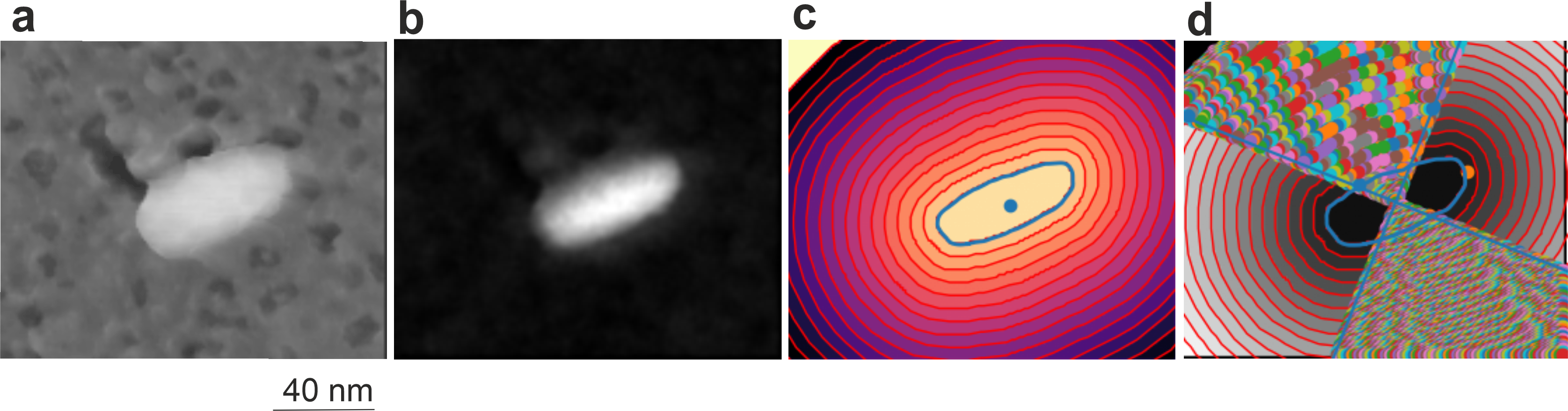}
    \caption{\textbf{EELS Map Segmentation} \textbf{(a)} HAADF image of disordered two phase metamaterials \textbf{(b)} identification of the Sn nanoparticle contour based on the bulk plasmon peak of Sn. \textbf{(c)} identification of perimeter coordinates and dividing up the maps into shells around the nanoparticles with defined thickness of 5 pixel and \textbf{(d)} averaging the spectra in segments and shells with a chosen opening angle here 45°}
\label{fig:segmentation}
\end{figure*}
To discern between inhomogeneous field distribution we divide our EELS map around the Sn NPs into quadrants with an opening angle of 45° towards the dipole axis. For this we developed a series of python scripts for Gatan. The scripts first identify the particle based on the Sn bulk plasmon peak (14eV), shown in Figure \ref{fig:segmentation} b. Then the coordinates of the center as well as the perimeter coordinates and the semiaxis orientations are determined. After this the maps are divided up into shells with a chosen thickness 5 nm (particle 1) respectivly 6 nm (particle 2), which corresponds to a thickness of 5 pixels in the EELS maps, around the NPs  (Figure \ref{fig:segmentation} c). Finally the EELS spectra are averaged in the shell segments with a chosen opening angle (Figure \ref{fig:segmentation} d). 

\subsection{Mechanical Model Approximation}
\begin{figure*}[h!]
  \includegraphics[width=1\textwidth]{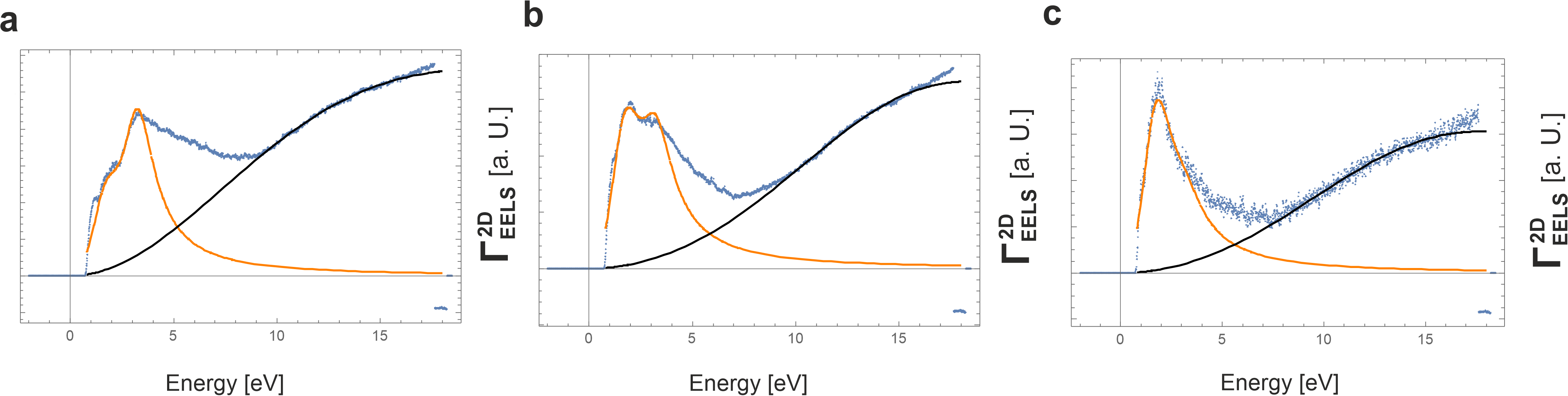}
    \caption{\textbf{Fit of Coupled oscillator model to experimental EELS data} (a),(b) and (c) The blue lines represent the EELS data of particle 1 in the direction where the dipolar Sn localized surface plasmon mode is dominant in shells 3, 10 and 16 respectively. The black line is the fit of the bulk plasmon peak contribution, which consequently has been removed prior to fitting for $\Omega$ and the orange line the fitted mechanical oscillator solution.}
\label{fig:oscilatorfit}
\end{figure*}

Two driven coupled oscillators  located at positions $r_i$ and driven by the force $F(r_i,t)$ in the quasi-static limit are represented by the effective Hamiltonian
\begin{equation}
\scriptsize
    H= \sum_{i=1} ^{N} \frac{P_i^2}{2 m_i} + \frac{1}{2}m_i \omega_i^2 Q_i^2 -\sum_{i\neq j} Q_i \Omega_{ij} Q_j
    - \sum_{i=1} ^{N} F(r_i,t) Q_i,
\end{equation}
where $Q_{i}$ and $P_{i}$ represent the displacement coordinate and conjugate momentum, $m_i$ the effective mass, and $\omega_i$ the resonance frequency of the $i$th dipolar mode. Furthermore, $\Omega_{ij}$ is the coupling strength for pairwise near-field dipole-dipole interaction.  By limiting our system to a mechanical two particle model, we exclude the interaction of the GNMs with each other for simplification purposes and only focus on the interaction of the GNMs with the dipolar mode of the Sn NPs. 
\\
\\
We first assume two non interacting oscillators, approximating the outermost shell as wells as the first shell surrounding the NP, by setting N = 1 respectively, resulting in 
\begin{equation}
    \ddot{Q} + \gamma \dot{Q} + \omega^2 Q = \frac{F(t)}{m}
\end{equation}
with an additional term $\gamma\dot{Q}$ accounting for damping. The average absorbed power can then be calculated by \begin{equation}
    <P_{abs}(t)>_T = \frac{1}{T} \int_{- T/2}^{- T/2} dt F(t) \dot{Q}(t).
\end{equation} 
We use this to seperatly fit the GNM and the dipolar mode of te Sn NP as two non-interacting harmonic oscilators. 
\\
\\
We then continue to introduce coupling between the two different systems, where the equation of motions are given by
\begin{equation}
    \ddot{x_1} + \gamma_1 \dot{x_1} + \omega_1 x_1 + \Omega x_2 = F_1
\end{equation} and
\begin{equation}
   \ddot{x_2} + \gamma_1 \dot{x_2} + \omega_2 x_2 + \Omega x_1 = F_2
\end{equation}
\\ 
From this the average absorbed power can then be calculated by
\\
\begin{equation}
 <P_{1c}> =  \omega Im[\frac{F_2 \Omega^2 + F_1 (i \gamma_2 \omega + \omega^2 - \omega_2^2)}{F1 ((-i \gamma_1 \omega - \omega^2 + \omega_1^2)(i \gamma_2 \omega +\omega^2 - \omega_2^2) + 2 \Omega^2}]
\end{equation}
and
\begin{equation}
 <P_{2c}> =    \omega Im[\frac{F_2 (-i \gamma_1 \omega - \omega^2 + \omega_1^2) - F_1 \Omega^2)}{F_1 ((-i \gamma_1 \omega - \omega^2 + \omega_1^2)(i \gamma_2 \omega +\omega^2 - \omega_2^2) + 2 \Omega^2}]
\end{equation}
\\
\\
This results in the two solutions $ <P_{1c}> +  <P_{2c}>$ respectively $ <P_{1c}> -  <P_{2c}>$.
\\
\\
Prior to fitting the experimental EELS data, we first remove the bulk plasmon peak, in the spectra, through a simple fit of the bulk plasmon peak by a forced and driven harmonic oscillator. We do this for every shell surrounding the nanoparticles.
\\
\\
Finally, we fit our coupled oscillator model to the EELS data and determine thereby $\Omega$ the coupling constant. The resulting fitted spectra for selected shells are shown in Figure \ref{fig:oscilatorfit}.
\subsection{Mode Coupling in zone 2}
\begin{figure*}[h!]
  \includegraphics[width=0.9\textwidth]{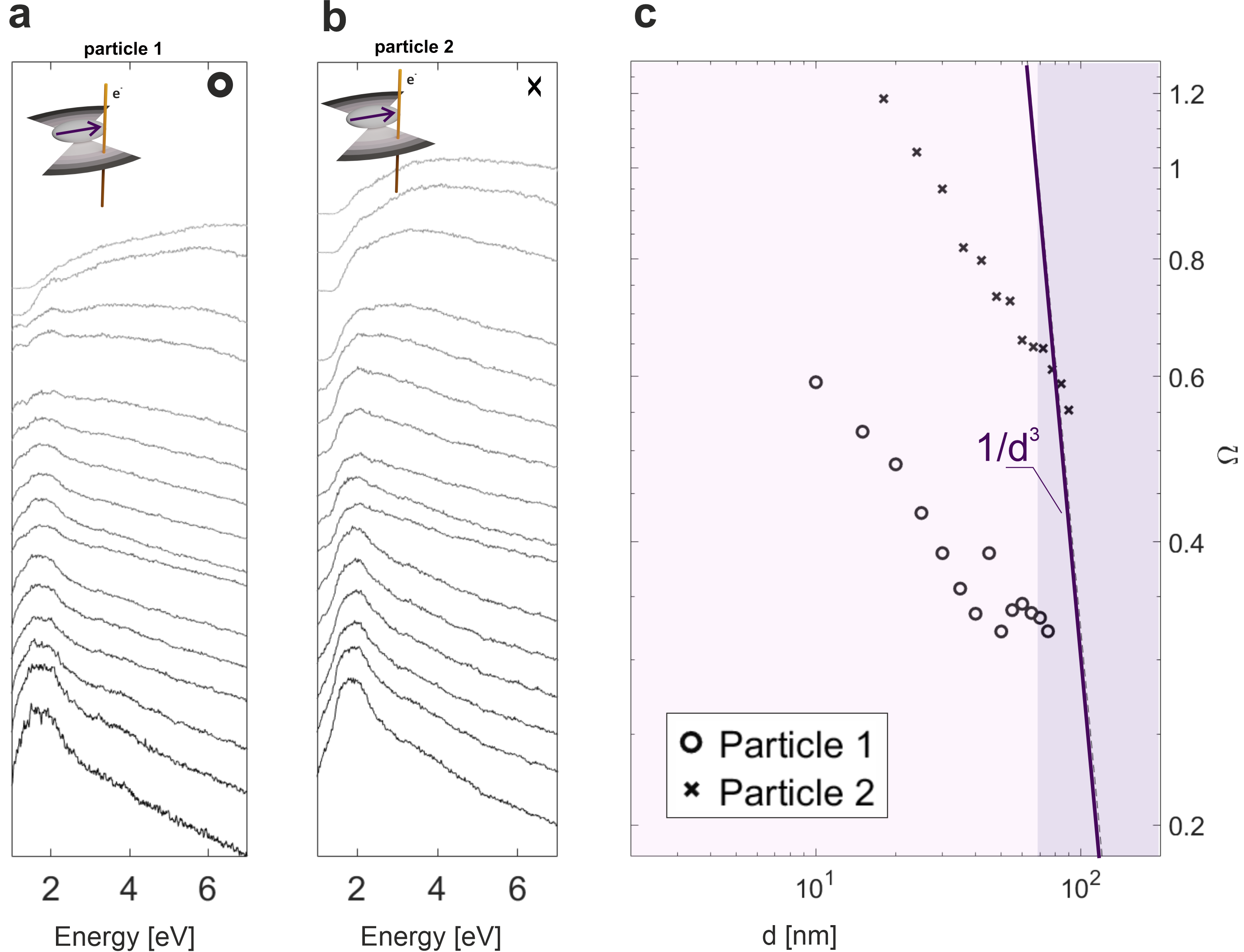}
    \caption{\textbf{Mode Coupling} \textbf{(a)} and \textbf{(b)} waterfall plot of EEL spectra averaged in the shell sectors around nanoparticle 1 and 2, where the higher order modes of the Sn NP are observed (Figure 3a II and III). The color transition from grey to black, corresponds to an increased inner radius of the sector shell around the NPs. The figure inset shows schematically the division of the EELS maps into sector shells around the NP with the arrow pointing in the dipolar direction \textbf{(c)} fitted coupling constant $\Omega$ for particle 1 (black circle) and particle 2 (black cross) as a function of the distance of the fitted mode to the Sn NP. The two different coupling regimes are shaded in purple, weak coupling with $\Omega$ scaling with $\frac{1}{d^3}$ and pink strong coupling in closer vicinity of the nanoparticle}
\label{fig:modecouplingparticle2}
\end{figure*}

In Figure \ref{fig:modecouplingparticle2} a and b waterfall plots of the EELS spectra for particle 1 and 2 are shown of the sector shell directions where higher order modes are dominant. Also in this direction, modulations of the spectra depending on the distance $d$ between the modes are observed.   
\par 
Like the other direction, the mechanical model was used to determine the coupling constant $\Omega$ as a function of distance, visible in \ref{fig:modecouplingparticle2} c. Again far away from the particle, $\Omega$ scales with $1/d^3$, indicating weak coupling. Moving closer to the particle, $\Omega$ stops following $1/d^3$, changing much faster due to interaction with the higher order modes of the Sn NP~\cite{interparticlecoupling}. 
\\
In this direction however, higher order modes start to become dominant faster, meaning the weak coupling regime is limited to further distances and the strong coupling region is reached faster.
\newpage
\bibliography{Paper_S.bib}